\documentclass[twocolumn,prb, superscriptaddress,showpacs]{revtex4-1}
\usepackage{amsmath}
\usepackage{graphicx}
\usepackage{multirow}
\bibpunct{[}{]}{,}{n}{}{}

\begin{document}

\title{Electronic coupling between Bi nanolines and the Si(001) substrate: An experimental and theoretical study}

\author{M. Longobardi}
\affiliation{Department of Quantum Matter Physics, University of Geneva, 24 Quai Ernest-Ansermet, CH-1211 Geneva 4, Switzerland}
\author{C. J. Kirkham}
\affiliation{National Institute for Materials Science (NIMS), Namiki 1-1, Tsukuba, Ibaraki, 305-0044, Japan}
\affiliation{London Centre for Nanotechnology and Department of Physics and Astronomy,University College London, London WC1E 6BT, United Kingdom}
\author{R. Villarreal}
\affiliation{Department of Quantum Matter Physics, University of Geneva, 24 Quai Ernest-Ansermet, CH-1211 Geneva 4, Switzerland}
\author{S. A. K\"{o}ster}
\affiliation{Department of Quantum Matter Physics, University of Geneva, 24 Quai Ernest-Ansermet, CH-1211 Geneva 4, Switzerland}
\author{D. R. Bowler}
\affiliation{London Centre for Nanotechnology and Department of Physics and Astronomy,University College London, London WC1E 6BT, United Kingdom}
\affiliation{International Center for Materials Nanoarchitectonics (MANA), National Institute for Materials Science (NIMS), Namiki 1-1, Tsukuba, Ibaraki, 305-0044, Japan}
\author{Ch. Renner}
\affiliation{Department of Quantum Matter Physics, University of Geneva, 24 Quai Ernest-Ansermet, CH-1211 Geneva 4, Switzerland}

\begin{abstract} 
Atomic nanolines are one dimensional systems realized by assembling many atoms on a substrate into long arrays. The electronic properties of the nanolines depend on those of the substrate. Here, we demonstrate that to fully understand the electronic properties of Bi nanolines on clean Si(001) several different contributions must be accounted for. Scanning tunneling microscopy reveals a variety of different patterns along the nanolines as the imaging bias is varied. We observe an electronic phase shift of the Bi dimers, associated with imaging atomic p-orbitals, and an electronic coupling between the Bi nanoline and neighbouring Si dimers, which influences the appearance of both. Understanding the interplay between the Bi nanolines and Si substrate could open a novel route to modifying the electronic properties of the nanolines.
\end{abstract}

\maketitle

\section{Introduction}

\begin{figure}[ht!]
\centering
\includegraphics[width=1\columnwidth]{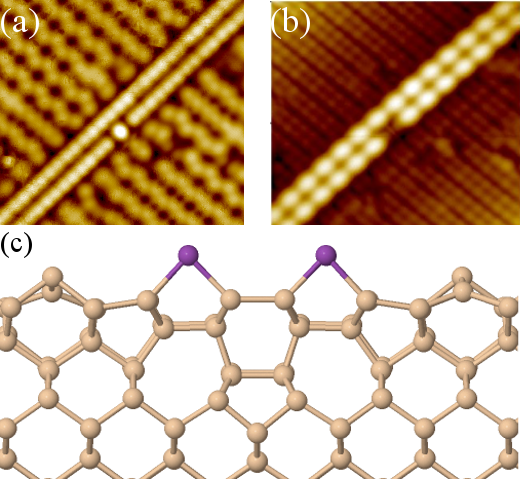}
\caption{(Color online) Appearance and structure of the Bi nanoline. $6.0\times5.5$~nm$^2$ experimental STM micrographs of the same Bi nanoline at (a) $-2.0$~V (0.1~nA) and (b) $+2.0$~V (0.1~nA). (c) Ball and stick model of the Bi nanoline, showing the subsurface Haiku structure. Si atoms are in beige (light) and Bi in purple (dark).}
\label{fig:Haiku}
\end{figure}

Self-assembled nanolines on surfaces have attracted considerable attention in recent decades because their low dimensional architectures offer the possibility to explore the exotic physics which emerges in one dimension (1D). While strong hybridization is expected for adatoms on metallic surfaces, semiconductors can offer a less electronically coupled environment~\cite{Schranz1998, Folsch2009}. With this in mind, many different self-assembled systems have been realized on semiconductor surfaces, such as Au chains stabilized on Si(335)~\cite{Au_on_Si(335)} and Si(553)~\cite{Au_on_Si(553)_magnet,Krawiec2010}, Pb wires on Si(557)~\cite{Pb_on_Si(557)}, and Au-induced wires on Ge(001)~\cite{Schafer}. In this context, self-assembled Bi nanolines on Si(001)~\cite{Miki1999,Bowler2000,Miwa,Owen2006} represent a special system, as shown in Fig.~\ref{fig:Haiku}(a) and (b). They do not need a stepped surface to grow, their length is limited only by the size of the Si terraces, and they can reach micrometer lengths, without kinks.

The Bi nanolines have been studied in detail and their structural properties are well established, including the striking 5-7-5 membered rings of the Haiku structure shown in Fig.~\ref{fig:Haiku}(c). The Haiku structure can be exposed by hydrogenation of the nanolines~\cite{Bianco2011}, and offers an attractive template for 1D Si dangling bond (DB) structures~\cite{Bianco2013}, or the self-assembly of other nanolines. However, the electronic properties of the nanolines, and in particular the influence of the Si substrate, are still not precisely known and understood~\cite{Belosludov2007,Javorsky2010}.

In this paper we investigate the impact of the Si substrate on the electronic structure of Bi nanolines on clean Si(001) by means of scanning tunneling microscopy (STM) and spectroscopy (STS) combined with density functional theory (DFT) simulations. We analyze the bias dependence of STM micrographs over a wide energy range, and focus on the little explored region close to the Fermi level (low bias). By comparison of STM micrographs and DFT simulations, we demonstrate that the local configuration of the Si substrate plays a key role in determining the electronic structure of the nanolines.

\section{Methods}
\subsection{Experiment}

All sample preparations and measurements were carried out in an ultrahigh vacuum (UHV) LT-STM Omicron system with base pressure better than $2\times 10^{-11}$~mbar and equipped with $in~situ$ facilities for sample heating and Bi evaporation. Si substrates were cut from commercial polished p-type Si(001) wafers (B doped, 0.008-0.015~$\Omega$cm), and chemically etched shortly before introducing them into the UHV chamber. Atomically flat Si(001) surfaces were routinely obtained after direct current outgassing at $700^\circ$C for 12~h, followed by repeated flashing at up to $1200^\circ$C, with the base pressure of the UHV chamber kept below $2\times 10^{-9}$~mbar. The quality of the Si(001) surface was controlled by monitoring in real time the reflected high-energy electron diffraction (RHEED) pattern during the heating process, and by STM imaging. Bismuth was evaporated from a ceramic crucible onto the substrate kept at around $500^\circ$C (checked by an optical pyrometer). Following Bi deposition, the sample was annealed at the same temperature for 4~min. The coverage was controlled by adjusting the flux, deposition time, and annealing time. A characteristic arc connecting the Si diffraction spots in the RHEED pattern was observed during the formation of the Bi nanolines. The arc typically appeared after several minutes of Bi deposition, becoming thicker and brighter during the annealing. The sample was subsequently transferred to the STM chamber. All STM images were taken in constant current mode at 77~K, with the bias voltage applied to the sample. STM tips were either made of mechanically cut PtIr wires or electrochemically etched W wires, with no observable differences in the data.

\subsection{Theory}

DFT~\cite{DFT1,DFT2} as implemented in the Vienna $ab~initio$ simulation package (VASP)~\cite{Vasp1,vasp2} version 4.6.34  was used to perform the calculations. Ultrasoft pseudopotentials (US-PPs)~\cite{Vanderbilt1990} with the PW91 exchange-correlation functional~\cite{Perdew1992} were used. For Bi, the $6s^{2} 6p^{3}$ electrons were treated as valence, the rest as core. The experimental bulk Si lattice constant of $a_{0} = 5.4306$~\AA~was used throughout.

The Si(001) substrate was represented by a periodically repeated ten layer slab model, with a reconstructed surface layer consisting of buckled Si dimers in the $p(2\times2)$ configuration. We also considered $c(4\times2)$ and mixed configurations. The Si surface consisted of two rows of Si dimers, each ten dimers long, with six regular dimers and a reconstructed Haiku region spanning the remaining four. Depending on the surface dimer arrangement, an eleven dimer long cell was sometimes used to allow for proper symmetry matching at the boundaries. The Bi nanoline itself consisted of a pair of Bi dimers per row, atop the Haiku region. As an extension to this model, we also considered various H coverages for the surface Si dimers. The bottom Si layer was terminated by H atoms in a dihydride structure, with both the H atoms and the bottom two Si layers held fixed, in order to simulate a bulk-like environment. All remaining atoms were allowed to move. Periodic images of the surface were separated by a vacuum gap of 12.73~\AA~to prevent interaction between repeated images.

An energy cut off of 250~eV was used, with a $(2\times1\times1)$ Monkhorst-Pack $k$-mesh along the direction of the Bi nanoline. A denser $k$-mesh, of up to $(10\times1\times1)$, did not affect our conclusions. Inclusion of the $\Gamma$ $k$-point improved the match to experiment at simulated biases around 1.0~V. Geometry optimizations were performed with a 0.02 eV/\AA~convergence condition for the forces on each atom. All calculations were spin polarised with no restrictions placed on the spins.

Simulated STM images were produced using the Tersoff-Hamann method, as implemented in BSKAN33~\cite{Hofer2003}, for biases between $\pm$3.0~V. Intervals of 0.1~V were sufficient to capture most details of the STM appearance, except for 0.8--1.2~V where narrower intervals of 0.01~V were used.

\section{Results and Discussion}
\subsection{Scanning tunneling spectroscopy features}

\begin{figure}[ht!]
\centering
\includegraphics[width=1\columnwidth]{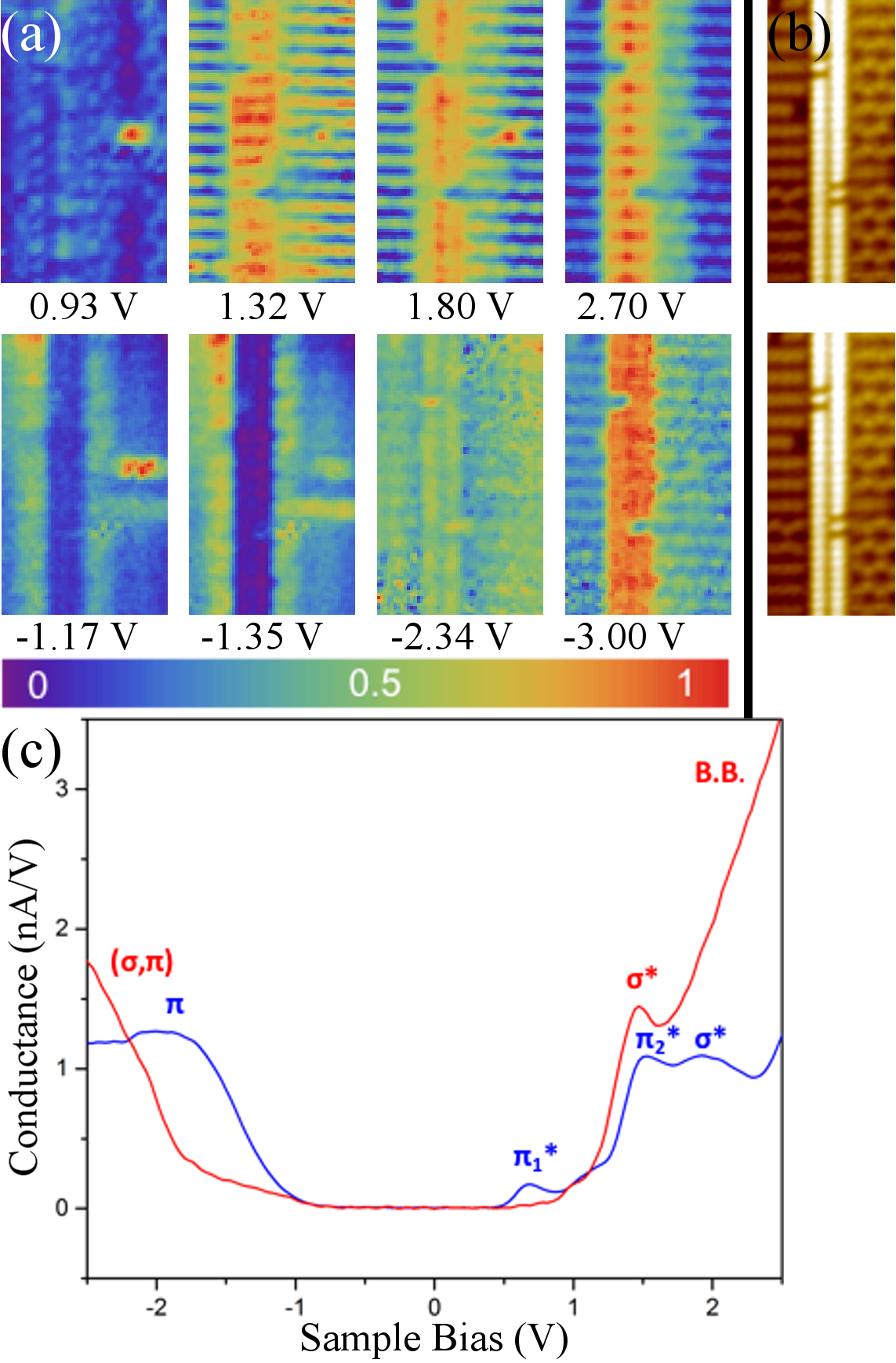}
\caption{(Color online) (a) dI/dV maps of a Bi nanoline at different bias voltages. (b) Corresponding STM micrograph of the same location at $-3.0$~V (0.2~nA), with images aligned to defects in the nanoline (same image shown twice). (c) Averaged dI/dV spectra measured on the Bi nanoline (red) and on the bare Si (blue).}
\label{fig:STS}
\end{figure}

We explore the electronic landscape of the Bi nanolines by STS mapping of the local electronic density of states with atomic-scale resolution. Fig.~\ref{fig:STS} shows dI/dV maps of a Bi nanoline at various voltages, and a characteristic tunneling spectrum, compared against the Si surface.

The electronic structure of the Si(001) surface is well established~\cite{Okada,Hata1999,Wolkow} and all of the dI/dV peaks observed in Fig.~\ref{fig:STS}(c) can be assigned to backbond and surface states~\cite{Nakayama2006,Owen2006}. The peak around 2.0~V reflects the $\sigma^{*}$ Si-Si bond. The peaks at 1.4 and 0.7~V reflect the buckling induced splitting of the $\pi$ state into $\pi_2^{*}$ and  $\pi_1^{*}$ states, respectively, and the shoulder at $-1.8$~V corresponds to the $\pi_1$ state.

By analogy, the dI/dV spectrum of the Bi nanoline can also be described in terms of $\pi$-type states and bulk states. Unlike the Si surface, the Bi nanoline consists of flat Bi dimers, without DBs. In addition, it is known that the Haiku structure induces short-range stress in the surrounding Si surface, straining neighbouring Si dimers and modifying their electronic properties~\cite{Owen2010}. The presence of the Bi nanolines replaces the $\pi$-type states of the bare Si spectra with the $\sigma^{*}$ state around 1.4~V. Below this state, around 1.0~V, a broad shoulder can also be detected. Since the $\pi^{*}$ and $\pi$ states of the Bi dimers are both located below the valence band, we cannot assign this shoulder to Bi atoms~\cite{Mark2005}. The backbond Bi-Si states at higher positive bias are deeply shifted towards the bulk conduction band.

In the dI/dV maps, at very low negative bias, around $-1.3$ to $-1.0$~V,  the Si dimers adjacent to the Bi nanolines appear brighter than the Si background [Fig.~\ref{fig:STS}(a)]. This is a known characteristic feature of the expected short-range stress. Interestingly, the nanolines appear darker than the rest of the surface, with the exception of a narrow bright line along the center. We ascribe these states to traces of the Haiku states that lie in the middle of the Haiku core~\cite{Bianco2011}.

\subsection{Scanning tunneling microscopy features}

To gain further insight into the electronic structure of the Bi nanolines, we study their bias dependent STM appearance, aided by DFT simulations to explore biases and structures both within and beyond the reach of experiment. We place particular emphasis on the little explored low bias regime, close to the Fermi level (E$_F$). When comparing simulations and experiment, it should be noted that equivalent simulated biases are consistently about 0.5~V lower than in experiment, due to the well-known understimate of gaps by DFT.

Features at high positive and negative bias are already well understood~\cite{Belosludov2007,Javorsky2010}, so will not be recounted in detail here. The individual Bi atoms of the nanoline are nicely resolved in high negative bias images [Fig.~\ref{fig:Haiku}(a)], whereas at high positive bias only complete dimers are imaged as a single broad bright spot [Fig.~\ref{fig:Haiku}(b)]. A more detailed analysis and comparison to calculations is provided in the Supplemental Material~\cite{Supplemental}.

\begin{figure}[ht!]
\centering
\includegraphics[width=1.0\linewidth]{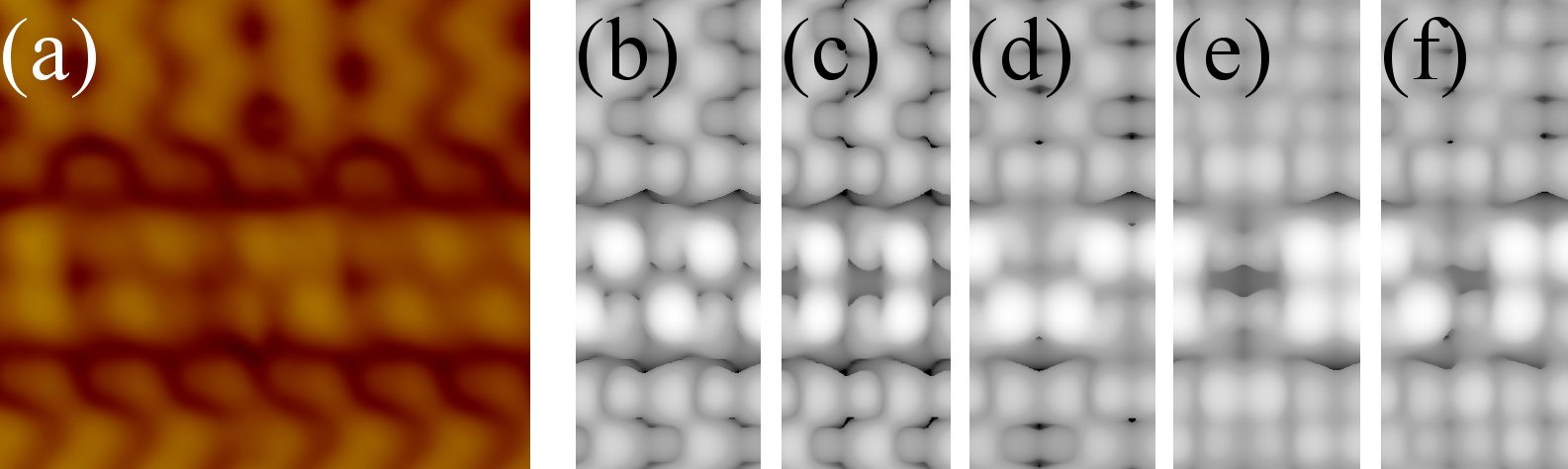}
\caption{(Color online) (a) $3.5\times3.1$~nm$^2$ experimental STM micrograph of a Bi nanoline at 1.1~V (0.2~nA), compared against simulated STM images of the Bi nanoline at 0.7~V for various different Si surface arrangements. (b) $p(2\times2)$, (c) $p(2\times2)$ with a mirror plane through the center of the nanoline, (d) $c(4\times2)$, (e) $c(4\times2)$ with a mirror plane, (e) $p(2\times2)$ and $c(4\times2)$.}
\label{fig:lowbias}
\end{figure}

The appearance of the Bi nanolines is most remarkably modified at positive biases below 1.5~V. Below 1.1~V, the Bi nanolines exhibit very intriguing patterns, strikingly similar to the bare Si surface, as shown in Fig.~\ref{fig:lowbias}(a). This is surprising since the Bi dimers are structurally all identical and flat. Inspired by this resemblance, we simulated STM images for different surface reconstructions, including various combinations of $p(2\times2)$ and $c(4\times2)$, as shown in Fig.~\ref{fig:lowbias}(b)--(f). In each case, the appearance of the Bi nanoline changes based on the Si reconstruction, resulting in a variety of patterns which perfectly match the experimentally observed diversity along the Bi nanoline. Since there is no structural difference between the Bi nanoline in each case, this must be a purely electronic effect. In all of these images, the bright spots on the Bi nanoline appear next to neighboring down buckled Si atoms, suggesting an electronic coupling between these atoms.

\begin{figure}[ht!]
\centering
\includegraphics[width=1\columnwidth]{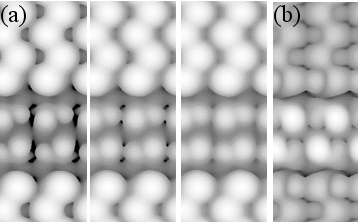}
\caption{Simulated STM of the Bi nanoline, with a $p(2\times2)$ Si reconstruction, showing regions inaccessible to experiment. (a) $-0.3$ to $-0.5$~V (left to right). (b) +0.7~V, provided for comparison. At negative bias the bright spots on the Bi nanoline neighbor up buckled Si atoms, at positive bias they neighbor down buckled Si atoms.}
\label{fig:lowbias2}
\end{figure}
 
Similar behavior is observed in simulations at low negative biases, shown in Fig.~\ref{fig:lowbias2}(a), which are inaccessible to experiment. A zigzag pattern is observed, but far dimmer than the positive bias pattern in Fig.~\ref{fig:lowbias2}(b). In addition, the slightly brighter spots that make up the zigzag appear next to up buckled Si atoms, rather than down buckled ones. Therefore, electronic coupling exists between the Bi nanoline and the substrate at both positive and negative biases, but the effect is stronger at positive biases.

\begin{figure}[ht!]
\centering
\includegraphics[width=1\columnwidth]{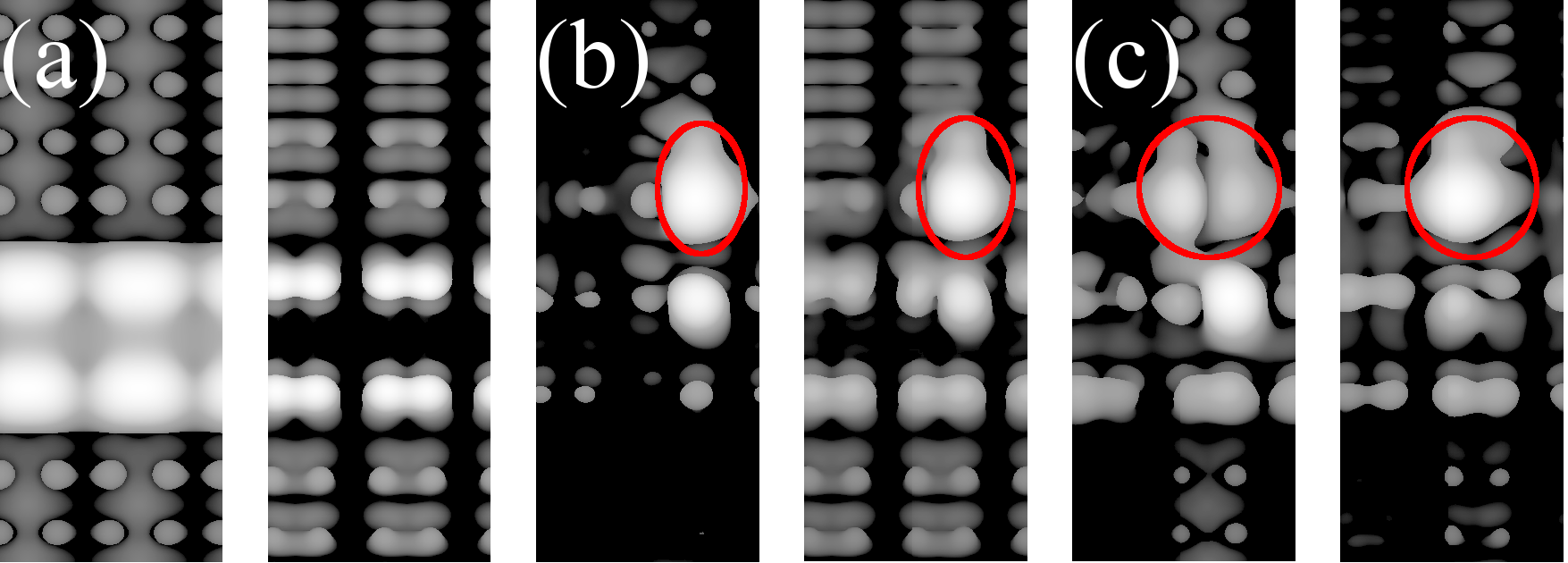}
\caption{(Color online) Simulated STM images for Bi nanolines with surface H passivation, at positive (left) and negative (right) bias. (a) Full H passivation at 0.8 and $-0.5$~V, (b) one DB at 0.7 and $-0.5$~V, (c) buckled clean Si dimer at 0.6 and $-0.5$~V.  In (c) the up buckled Si is on the left. Circles indicate depassivated regions. The link between a buckled Si substrate and the charge structure along the Bi nanoline is obvious.}
\label{fig:HPassivation}
\end{figure}

To clarify the details of the Bi nanoline--Si substrate coupling we performed further calculations, using selective surface H passivation to isolate individual Si DBs, as shown in Fig.~\ref{fig:HPassivation}. For the fully H passivated surface [Fig.~\ref{fig:HPassivation}(a)] the first occupied and unoccupied Bi states are observed at $-0.5$  and 0.8~V, respectively. Unlike the clean surface, patterns of bright spots are not observed between these biases, due to the removal of the Si DB states. Removing a single H from the nearest Si dimer [Fig.~\ref{fig:HPassivation}(b)] immediately yields a bright spot next to the Si DB, both in the positive and negative bias simulated Bi nanoline images. No effect was observed for DBs on more distant Si dimers, meaning the coupling is a short range effect. Removing two H to form a buckled Si dimer next to the Bi nanoline [Fig.~\ref{fig:HPassivation}(c)] reproduces the behavior observed on the clean Si surface shown in Fig.~\ref{fig:lowbias2}. At positive bias a bright spot appears on the Bi nanoline next to the down buckled Si, whereas at negative bias the brighter spot appears next to the up buckled Si. The model thus implies that the low bias appearance is due to an electronic coupling between the highest occupied or lowest unoccupied Bi dimer states and the neighboring Si DB states.

In this regard, the Bi nanolines represent a very intriguing system, since their electronic properties could be altered by simply changing the reconstruction of the Si background, which could be triggered by temperature changes~\cite{wolkow92}, manipulated by a scanning probe tip~\cite{cho96,yoshida04,sweetman11}, or by hydrogenation, as demonstrated by our simulations above. Bi nanolines themselves are very stable against changes in temperature (up to 750~K), whereas the Si surface is very sensitive, and experiences a phase transition from $p(2\times2)$ to $c(4\times2)$ at low temperatures. Simply by heating or cooling the sample, we could provide a local modification of the Si background, and thus an alteration of the electronic properties of the Bi nanolines.

\begin{figure}[ht!]
\centering
\includegraphics[width=1\columnwidth]{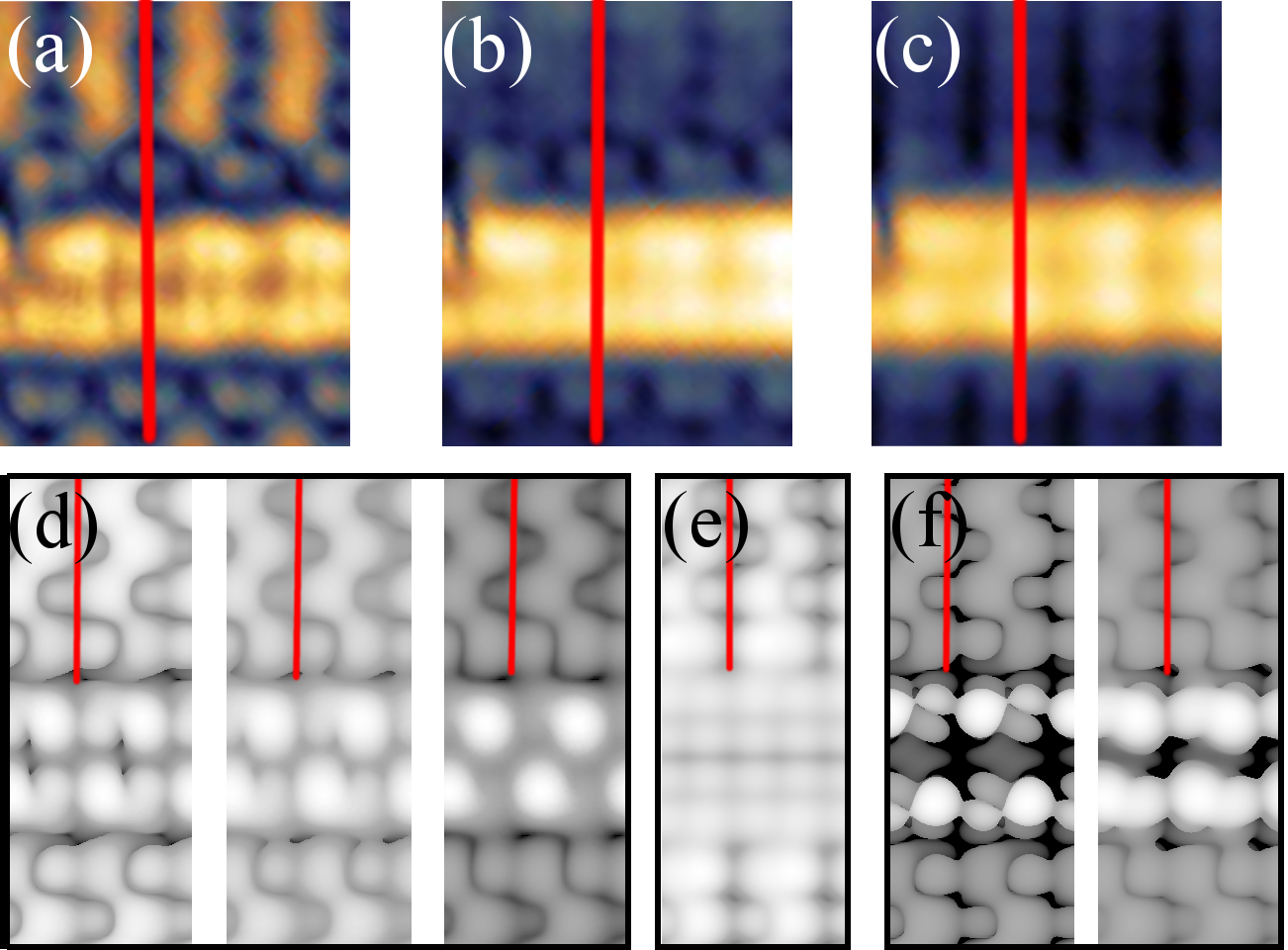}
\caption{(Color online) $2.4\times3.1$~nm$^2$ STM micrographs of a Bi nanoline at (a) 1.20~V, (b) 1.40~V and (c) 1.45~V. 
DFT simulations of (d) the clean Bi nanoline at 0.3, 0.5 and 0.7~V, (e) the bare Haiku (no Bi) at 0.7~V,
and (f) of a Bi dimer line without the Haiku structure at 0.4 and 0.6~V. In all case, red lines indicate the position of the trench in the Si background.}
\label{fig:Coupling}
\end{figure}

The electronic coupling between the Bi nanoline and the neighboring Si dimers also has a subtle effect on the electronic properties of the Si dimers. It is well known that the Si surface undergoes a phase shift around 1.4~V, with the dark stripes in STM images shifting from the trench between dimer rows to the center of the dimer rows~\cite{Hata1999,Okada2001}. We confirm this behavior of the Si background in Fig.~\ref{fig:Coupling}(a), except for the neighboring Si dimers, which appear out of phase with, and brighter than, the rest of the surface below 1.45~V. This suggests the neighboring Si dimers experience a phase shift at an even lower bias than the background. A similar effect has been observed for other atomic wire systems, including Mn~\cite{RennerMn2015} and In~\cite{Dong2001}. For Bi nanolines, this has previously been attributed to strain effects from the Haiku structure~\cite{Owen2003}.

To examine this idea, and explore when the neighboring Si experiences a phase shift, we simulate STM images for three different systems, namely the regular Bi nanoline [Fig.~\ref{fig:Coupling}(d)], the bare Haiku structure (without Bi) [Fig.~\ref{fig:Coupling}(e)], and a Bi nanoline without the underlying Haiku structure [Fig.~\ref{fig:Coupling}(f)]. For the regular Bi nanoline the neighboring Si dimers experience a phase shift at a far lower bias than the background, with the first signs of a phase shift around 0.3~V, becoming very clear at 0.7~V. If this was purely a strain effect from the Haiku structure, we would expect the same behavior in the absence of Bi, but this is not the case, as can be seen in Fig.~\ref{fig:Coupling}(e). The neighboring Si no longer experience a phase shift at a lower bias, keeping in phase with the Si background at all biases. However, this Si does appear brighter than the background Si. To test if both Bi and the Haiku structure were necessary, we considered two lines of Bi dimers adsorbed above regular Si [Fig.~\ref{fig:Coupling}(f)]. Here the adjacent Si dimers experience a phase shift at 0.6~V, still before the Si background. Therefore, we conclude that electronic coupling between the Bi dimers and the neighboring Si induces a phase shift at a very low bias, outside of our experimental range. The Haiku induced strain effect accounts for the brightness difference, but is not a necessary component of the phase shift.

\begin{figure}[ht!]
\centering
\includegraphics[width=1.0\columnwidth]{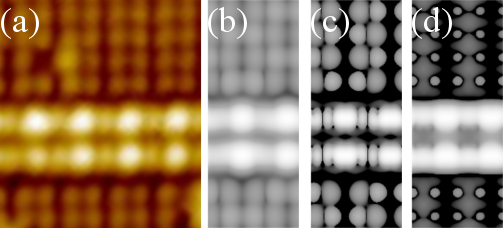}
\caption{(Color online) (a) $2.6\times3.1$~nm$^2$ experimental STM micrograph of a Bi nanoline at 1.5~V (1~nA), compared against simulated STM for a Bi nanoline on $c(4\times2)$ Si at 1.1~V for isosurface values of (b) 0.1 and (c) 1 arbitrary units. (d) Simulated STM for a Bi nanoline on H:Si(001) at 0.88~V.}
\label{fig:Intermediate}
\end{figure}

The appearance of the nanolines changes again around 1.5~V, where it splits into brighter and dimmer spots, as shown in Fig.~\ref{fig:Intermediate}(a). The brighter spots correspond to the region between the Bi dimers, and the dimmer spots to the dimers themselves. Recall that the Si background also experiences a phase shift, so this effect can be easily missed at first glance. This effect is limited to a narrow bias range of less than 0.2~V, and quickly changes to the high bias appearance shown earlier in Fig.~\ref{fig:Haiku}(b). We reproduced this behavior in simulated STM images, shown in Fig.~\ref{fig:Intermediate}(b), with the match to experiment improved when the $\Gamma$ $k$-point was included. For comparison, results without the $\Gamma$ $k$-point are provided in the Supplemental Material~\cite{Supplemental}. We resolved further details of the electronic states responsible for this appearance by changing the imaging settings [Fig.~\ref{fig:Intermediate}(c)]. The Bi dimers look remarkably like overlapping $p$-orbitals, with bonding orbitals between the dimers, and anti-bonding orbitals on the dimers. This was confirmed via an analytical model, which is included in the Supplemental Material~\cite{Supplemental}. Our simulations still show limited coupling to the Si substrate, with a slight asymmetry to the dimmer spots, matching to the earlier low bias patterns in Fig.~\ref{fig:lowbias}. If we remove the substrate coupling via H passivation, then the dimmer spots become clearer [Fig.~\ref{fig:Intermediate}(d)].

The excellent match between experiment and DFT suggests that it is possible to see isolated $p$-orbitals along the Bi nanoline, with the phase shift due to imaging a specific combination of bonding and anti-bonding orbitals. Further study of this bias range could prove of interest for fundamental physics, especially following H passivation of the Si surface.

\section{Conclusions}

In conclusion, we performed a detailed experimental and theoretical analysis of Bi nanolines on Si(001), demonstrating how the electronic structure of the Bi nanolines strongly depends on the local structure of the Si substrate. We found an electronic coupling between the Bi dimers of the nanoline, and the adjacent Si dimers. This coupling results in a variety of patterns along the Bi nanoline at low bias, and modifies the electronic properties of the adjacent Si dimers, causing them to experience an electronic phase shift at a lower bias than the Si background. Moreover, we found that the Haiku induced strain only contributes to the increased brightness of these dimers. Our analysis suggests that the electronic properties of Bi nanolines could be tuned by choosing the appropriate Si(001) reconstruction, which could be achieved through standard techniques such as chemical etching, $in~situ$ flashing, or manipulation with a scanning probe tip. In addition, we demonstrated the ability to image atomic $p$-orbitals within the Bi nanoline at specific biases, which could be of interest in the fundamental study of atomic orbitals.

\begin{acknowledgments}
We thank F. Bianco, H. Zandvliet and J. Owen for stimulating discussions. We thank G. Manfrini for his technical assistance in the STM laboratories. D.R.B. and C.J.K. acknowledge financial support from a UCL Impact Studentship; C.R., M.L., R.V and S.A.K. acknowledge financial support from the Swiss National Science Foundation.
\end{acknowledgments}

\bibliography{binanolines}

\end{document}